\documentclass[aps,pra,showpacs,twocolumn,groupedaddress]{revtex4-1}

\usepackage{graphicx}
\usepackage{dcolumn}
\usepackage{bm}
\usepackage{amsmath}
\usepackage{amssymb}
\usepackage{latexsym}
\usepackage{epsfig}
\usepackage{amsbsy}
\usepackage{array}
\usepackage{amssymb}
\usepackage{setspace}
\usepackage{bm}
\usepackage{epstopdf}
\usepackage{subfigure}

\def\sint{\ifmmode{- \!\!\!\!\!\! \int}
    \else{\hbox{$- \!\!\!\! \int \ $}}\fi}

\begin{document}

\title{Lyapunov-based quantum synchronization in a designed optomechanical system}

\author{Wenlin Li}
\author{Chong Li}
\email{lichong@dlut.edu.cn}
\author{Heshan Song}
\email{hssong@dlut.edu.cn}
\affiliation{School of Physics and Optoelectronic Engineering, Dalian University of Technology, 116024, China}

\begin{abstract}
We extend the concepts of quantum complete synchronization and phase synchronization, which are proposed firstly in [Phys. Rev. Lett, 111 103605 (2013)], to more widespread quantum generalized synchronization. The generalized synchronization can be considered as a necessary condition or a more flexible derivative of complete synchronization, and its criterion and synchronization measurement are further proposed and analyzed in this paper. As an example, we consider two typical generalized synchronizations in a designed optomechanical system. Unlike the effort to construct a special coupling synchronization system, we purposefully design extra control fields based on Lyapunov control theory. We find that the Lyapunov function can adapt to more flexible control objectives, which is more suitable for generalized synchronization control, and the control fields can be achieved simply with a time-variant voltage. Finally, the existence of quantum entanglement in different generalized synchronizations is also discussed. 

\end{abstract}
\pacs{42.50.Wk, 05.45.Xt, 05.45.Mt, 03.65.Ud}
\maketitle
\section{Introduction}  
Complete synchronization and phase synchronization between two continuous variable (CV) quantum systems were first studied by \citeauthor{S1} in mesoscopic optomechanical systems \cite{S1}, and they also made the forward-looking prediction  that the quantum synchronization is of  potential but important applications in quantum information processing (QIP). Subsequently, quantum synchronization is deeply discussed in variety of quantum systems, such as cavity quantum electrodynamics \cite{S2}, atomic ensembles \cite{S3,S4}, van der Pol (VdP) oscillators \cite{S2,S5,S6,S11}, Bose-Einstein condensation \cite{S7} ,superconducting circuit system \cite{S8}, and so on. In these works, quantum synchronization is further extended from CV system to finite dimensional Hilbert space corresponding to more excellent quantum properties and the quantum correlation is also analyzed quantificationally in those synchronous quantum system \cite{S2,S3,S5,S9}. In addition, quantum synchronization criteria \cite{S3,S10,S11,S12} and synchronization between nodes in quantum network are still hot topics in the field of quantum synchronization theory.

Generally, existing quantum synchronization schemes can be attributed to the concept of coupling synchronization, i.e., one subsystem of the synchronous systems plays the role of controller acting on the other subsystem \cite{S1,S2,S3,S4,S5,S6,S11,S10,S9}. A significant advantage of this kind of direct linking is its high maneuverability. However, it still remains some difficulties to achieve the better applications in QIP with quantum synchronization. For the weak coupling of the quantum level, it is difficult to eliminate the difference between systems if it is big enough, and in fact, it is often in this case. Fundamentally, too strong driving or pump fields will compel systems to the form of the forced synchronization, just like what they are realized in previous works \cite{S1,S2,S3,S9}. This deficiency of coupling synchronization is a severe limitation, which causes other types of synchronizations are hardly realized in addition to complete and phase synchronizations. For examples, antiphase synchronization and projective synchronization, which are also widely applied in the classical synchronization fields \cite{GSS1,GSS2,GSS3}, were rarely discussed in quantum systems. 

In traditional control theory, besides the coupling terms, there exists an external controller which is  imposed  on response system in order to provide a more outstanding control capability, implying that a designed controller can establish a more flexible relationship between two controlled subsystems \cite{Con}. It enlightens us to think about such problems: can a more generalized synchronization (like above mentioned antiphase and projective synchronizations) be extended and obtained in quantum domain? If it works, what kinds of the criterion and measurement are needed in this quantum generalized synchronization? And most importantly, how are the controllers designed in order to satisfy various requirements corresponding to different kinds of generalized synchronizations?
   
For responsing above questions, in this paper, we study the general properties of different synchronization forms and expand them into quantum  mechanics based on Mari's complete  synchronization theory. The criterion and measurement of the generalized synchronization are also proposed and they will be divided into two orders for calculating and analyzing conveniently. Instead of directly establishing interaction between two subsystems, here we utilize Lyapunov control theory which has exhibited comprehensive applications in target quantum state preparation and suppressing decoherence to design the external controller \cite{Con,QC1,QC2}. Although the Lyapunov function is constituted by expectation values, our results show that the quantum fluctuations can also be effectively subdued by the controller. In addition, the classical and quantum correlations are considered via calculating the Lyapunov exponent and Gaussian Negativity. In particular, we demonstrate that CV entanglement can exist in generalized synchronization, however, it will disappear if  generalized synchronization tends to complete  synchronization. This phenomenon  is consistent with Mari's and Ameri's conclusions about entanglement in complete  synchronization \cite{S1,S2}. Therefore, we believe existing quantum complete  synchronization can be included in our generalized synchronization theory.
 
We organize this paper as follows: in Sec. \ref{quantum Generalized synchronization}, we introduce the definition and the properties, especially measurement method, of quantum generalized synchronization. In Sec. \ref{Lyapunov-based synchronization in optomechanical systems}, we analyze the dynamics of an optomechanical system, and realize two kinds of representative generalized synchronization (constant error synchronization in Sec. \ref{Constant error synchronization} and time delay synchronization in Sec. \ref{Time delay synchronization}) respectively via designing appropriate control field based on Lyapunov function theory. The correlation in generalized synchronization is also discussed in Sec. \ref{correlation in Generalized synchronization} and a summary is finally given in Sec. \ref{Results and discussion}.

\section{quantum Generalized synchronization} 
\label{quantum Generalized synchronization} 
We begin this section with a brief introduction of generalized synchronization and its expansion in quantum domain. Considering two general classical dynamics systems whose evolutions satisfy following equations:
\begin{equation}
\begin{split}
\partial_{t}x_{1}(t)=F(x_1(t))+U_{c1}(x_1,x_2)+U_{e1}\\
\partial_{t}x_{2}(t)=F(x_2(t))+U_{c2}(x_1,x_2)+U_{e2}
\end{split}
\label{eq:sys}
\end{equation}
here $x_{1,2}(t)\in R^n$ are the state variables of two systems in time $t$. $U_{c1,2}$ are the mutual couplings between systems and correspondingly, $U_{e1,2}$ are the external controllers belonging to their respective system. If there are continuous mappings $h_{1}, h_{2}:R^n\rightarrow R^k$ and following synchronization condition in Eq. (\ref{eq:classc}) can be achieved when $t\rightarrow\infty$, then two systems depending on $h_{1}$, $h_{2}$, $x_{1}$ and $x_{2}$ will be of the consistent evolution. 
\begin{equation}
\begin{split}
\lim_{t\rightarrow\infty}\left|h_1(x_1(t))-h_2(x_2(t))\right|\rightarrow 0
\end{split}
\label{eq:classc}
\end{equation}
This controllable correlation is named as generalized synchronization, and it will degenerate to common complete synchronization or phase synchronization via selecting $h_{i}(x_{i})=x_{i}$ or $h_{i}(x_{i})=\arg(x_{i})$, respectively. Similarly with Mari's measurement $S_c(t):=\langle \hat{q}_{-}^2(t)+\hat{p}_{-}^2(t)\rangle^{-1}$ \cite{S1}, generalized synchronization can be extended from classical to quantum by considering conjugate quantities simultaneously, and the corresponding measurement can be defined as: 
\begin{equation}
\begin{split}
S_g(t):=&\langle \hat{q}_{g-}^2(t)+ \hat{p}_{g-}^2(t)\rangle^{-1},
\label{eq:Mari}
\end{split}
\end{equation}
where $\hat{q}_{g-}:=(h_1(\hat{q}_1)-h_2(\hat{q}_2))/\sqrt{2}$ and $\hat{p}_{g-}:=(h_1(\hat{p}_1)-h_2(\hat{p}_2))/\sqrt{2}$ are the quantized generalized error operators. 

Nevertheless, it is not easy to use Eq. (\ref{eq:Mari}) directly in a concrete model. In some cases, $h_{1,2}(q,p_{1,2})$ are not strict physical descriptions because they are actually superoperators. On the other hand, it could be difficult to calculate $S_{g}(t)$ in CV quantum systems. Therefore, in order to analyze quantum synchronization in CV mesoscopic systems, we adopt mean--field approximation to simplify Eq. (\ref{eq:Mari}). Then synchronization measurement can be divided into two parts: the first order criteria is to describe the consistency of expectation values:
\begin{equation}
\begin{split}
&\lim_{t\rightarrow\infty}\left|h_1(q_1(t))-h_2(q_2(t))\right|\rightarrow 0\\
&\lim_{t\rightarrow\infty}\left|h_1(p_1(t))-h_2(p_2(t))\right|\rightarrow 0
\label{eq:liwenclass}
\end{split}
\end{equation}
and the second order measurement is to determine following quantum fluctuations:
\begin{equation}
\begin{split}
S'_g(t):=&\langle \delta q_{g-}^2(t)+ \delta p_{g-}^2(t)\rangle^{-1}
\label{eq:liwen}
\end{split}
\end{equation}
where $o$ refers to $\langle o\rangle$ and $\delta o:=\hat{o}-o$ for $o\in \{{q}_{g-},{p}_{g-}\}$.  

Physical meaning of Eq. (\ref{eq:liwenclass}) and Eq. (\ref{eq:liwen}) are more definite to explain quantum synchronization, i.e., systems' expectation values are required to satisfy the ``classical" generalized synchronization conditions and the perturbation on synchronization behavior caused by quantum effect is squeezed as much as possible. To verify this, Eq. (\ref{eq:liwen}) will be equivalent to Eq. (\ref{eq:Mari}) if the first order criteria are satisfied. Conversely, if a designed external field can not only make the evolution of systems to realize ``classical" generalized synchronization conditions, but also increase corresponding second order measurement $S'_{g}$, then it can be seen as an appropriate control field for realizing quantum synchronization. This is the basic idea of designing the control field. 

In some particular models, if $h_1$ and $h_2$ are selected as flat mappings, Eq. (\ref{eq:liwen}) can be further simplified to measure fluctuation
\begin{equation}
\begin{split}
S'_g(t):=&\langle \delta q_{-}^2(t)+ \delta p_{-}^2(t)\rangle^{-1}
\label{eq:liwenll}
\end{split}
\end{equation}
Here ${q}_{-}:=({q}_1-{q}_2)/\sqrt{2}$ and ${p}_{-}:=({p}_1-{p}_2)/\sqrt{2}$. Compared with Eq. (\ref{eq:liwen}), Eq. (\ref{eq:liwenll}) is more easy to be obtained via the covariance matrix of system. 
\section{Lyapunov-based synchronization in optomechanical system}
\label{Lyapunov-based synchronization in optomechanical systems} 
\begin{figure}[]
\centering
\begin{minipage}{0.48\textwidth}
\centering
\includegraphics[width=3in]{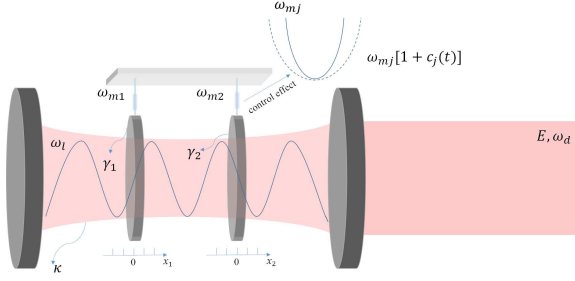}  
\end{minipage}
\caption{Diagram of optomechanical system corresponding to our model. Here two oscillators are placed at wave nodes of a Fabry–-P\' erot cavity and they couple with the cavity field via linear optomechanical interactions, and their origins are respectively set at the equilibrium positions.  
\label{fig:fig1}}
\end{figure}

We analyze Lyapunov-based synchronization in optomechanical system to more intuitively explain above theory of quantum generalized synchronization. Our model consists of two oscillators which couple with a Fabry–-P\' erot cavity together (see Fig. \ref{fig:fig1}). 
The Hamiltonian corresponding to this model can be divide into four parts: $H=H_{0}+H_{int}+H_{div}+H_c(t)$. Here $H_{0}=\omega_{l}a^{\dagger}a+\sum_{j=1,2}(\dfrac{w_{mj}}{2}\hat{p}_{j}^2+\dfrac{w_{mj}}{2}\hat{q}_{j}^2)$ is a sum of free Hamiltonians corresponding to the optical field and two oscillators. Moreover, $H_{int}=-g_{1}a^{\dagger}a\hat{q_{1}}-g_{2}a^{\dagger}a\hat{q_{2}}$ and $H_{div}=iE(a^{\dagger}e^{-i\omega_{d}t}-ae^{i\omega_{d}t})$ are the standard forms of optomechanical interaction and driving field respectively \cite{H}. $H_c(t)$ is an external control Hamiltonian which represents the coupling with  designed time-dependent fields. In our model, we consider such a form of control field which can create a deviation in respective potential term of two oscillators. This effect can be regarded as a time-dependent rescaling of the mirror frequency \cite{LE2}, i.e.,
\begin{equation}
\dfrac{w_{mj}}{2}\hat{q}_{j}^2\rightarrow\dfrac{w_{mj}}{2}[1+C_j(t)]\hat{q}_{j}^2
\label{eq:contorl}
\end{equation}
We will give a more detailed discussion about how to realize this form of control field using specific experiments in Sec. \ref{Results and discussion}. Then the Hamiltonian of the whole system is written as follow after a frame rotating:
\begin{equation}
\begin{split}
H=&\sum_{j=1,2}\{\dfrac{w_{mj}}{2}\hat{p}_{j}^2+\dfrac{w_{mj}}{2}[1+C_{j}(t)]\hat{q}_{j}^2-g_{j}a^{\dagger}a\hat{q_{j}}\}\\&-\Delta a^{\dagger}a+iE(a^{\dagger}-a)
\label{eq:Hamilton}
\end{split}
\end{equation}
In above expressions, $a$($a^{\dagger}$) is the annihilation (creation) operator for the optical field and correspondingly for $j=1,2$, $q_{j}$ and $p_{j}$ are dimensionless position and momentum operators of the oscillator $j$ respectively. $\Delta=\omega_d-\omega_l$ refers to the detuning between the frequencies of the laser drive and the cavity mode. $\omega_{mj}$ is the mechanical frequency. $g_{j}$ is the optomechanical coupling constant and $E$ is the drive intensity. In order to solve the dynamics of the system, we consider the dissipative effects in the Heisenberg picture and write the quantum Langevin equations as follows \cite{LE1,LE2,LE3}:
\begin{equation}
\begin{split}
&\partial_{t}a=(-\kappa+i\Delta)a+ig_{1}a\hat{q}_1+ig_{2}a\hat{q}_2+E+\sqrt{2\kappa}{a}^{in}\\ 
&\partial_{t}\hat{q}_j=\omega_{mj}\hat{p}_{j}\\
&\partial_{t}\hat{p}_j=-\omega_{mj}[1+C_j(t)]\hat{q}_{j}-\gamma_{j}\hat{p}_{j}+g_{j}a^{\dagger}a+\hat{\xi}_{j}
\label{eq:qle}
\end{split}
\end{equation}
Here $\kappa$ is the decay rate of optical cavity, and $\gamma_{j}$ is the mechanical damping
rate of each oscillator. ${a}^{in}$ is the input bath operator, which satisfies $\langle{a}^{in}{(t)}{a}^{in,{\dagger}}{(t')}\rangle=\delta(t-t')$ \cite{Noise0}. Similarly, $\hat{\xi}_{j}(t)$ is the Brownian noise operator describing the dissipative friction force acting on the $j$th mirror. In the Markovian approximation, the autocorrelation function of $\hat{\xi}_{j}(t)$ satisfies the relation: $\langle\hat{\xi}_j{(t)}\hat{\xi}_{j'}{(t')}+\hat{\xi}_{j'}{(t')}\hat{\xi}_j{(t)}\rangle/2=\gamma_{j}(2\bar{n}_b+1)\delta_{jj'}\delta(t-t')$, where $\bar{n}_b=[\exp ({\hbar\omega_j}/{k_BT})-1]^{-1}$ is the mean phonon number of the mechanical bath which gauges the temperature $T$ \cite{Noise1,Noise2,Noise3}. 

Solving directly a set of nonlinear differential operator equations like Eq. (\ref{eq:qle}) is  quite difficult, however, a mean--field approximation is acceptable in our mesoscopic  optomechanical model \cite{H,line1,line2,line3}. On the other hand, as we discussed in Sec. \ref{quantum Generalized synchronization}, the quantum synchronization measurement modifying by mean--field approximation can 
describe the generalized synchronization effect more accurately. Therefore, the every operator in Eq. (\ref{eq:qle}) can be rewritten respectively as a sum of its expectation value and a small fluctuation near the expectation value, that is, $a(t)=A(t)+\delta a(t)$, $\hat{o}(t)=o(t)+\delta o(t), o\in(q_{1,2},p_{1,2})$. After neglecting the high--order fluctuation terms, the ``classical" properties of our optomechanical system can be described by following nonlinear equations:
\begin{equation}
\begin{split}
&\partial_{t}A=(-\kappa+i\Delta)A+ig_{1}Aq_1+ig_{2}Aq_2+E\\ 
&\partial_{t}q_j=\omega_{mj}p_{j}\\
&\partial_{t}p_j=-\omega_{mj}[1+C_j(t)]q_{j}-\gamma_{i}p_{j}+g_{j}\vert A\vert^2
\label{eq:mean}
\end{split}
\end{equation}
and the corresponding quantum fluctuations can also be confirmed by:
\begin{equation}
\begin{split}
\partial_{t}\delta a=&(-\kappa+i\Delta)\delta a+\sum_{j=1,2} ig_{j}(q_{j}\delta a+A\delta q_{j})+\sqrt{2\kappa}{a}^{in}\\ 
\partial_{t}\delta q_j=&\omega_{mj}\delta p_{j}\\
\partial_{t}\delta p_j=&-\omega_{mj}[1+C_j(t)]\delta q_{j}-\gamma_{j}\delta {p}_{j}+g_{j}(A^{*}\delta{a}+A\delta a^{\dagger})+\hat{\xi}_{j}
\label{eq:fluctuations}
\end{split}
\end{equation}
Transforming the annihilation operators as the forms of $a=(\hat{x}+i\hat{y})/\sqrt{2}$ and $a^{in}=(\hat{x}^{in}+i\hat{y}^{in})/\sqrt{2}$ respectively. Then Eq. (\ref{eq:fluctuations}) can be rewritten more concisely as $\partial_{t}\hat{u}=S\hat{u}+\hat{\zeta}$ by setting the vectors $\hat{u}=(\delta x, \delta y, \delta q_1, \delta p_1, \delta q_2, \delta p_2)^{\top}$ and $\hat{\zeta}=(\hat{x}^{in}, \hat{y}^{in}, 0, \hat{\xi}_{1}, 0, \hat{\xi}_{2})^{\top}$, and the corresponding $S$ is a time-dependent coefficient matrix (see Appendix. A for more details). In this representation, the evolution of correlation matrix $D$ defined as 
\begin{equation}
\begin{split}
D_{ij}(t)=D_{ji}(t)=\dfrac{1}{2}\langle\hat{u}_i(t)\hat{u}_j(t)+\hat{u}_j(t)\hat{u}_i(t)\rangle
\label{eq:correlationmatrix}
\end{split}
\end{equation}
can be derived directly by (see \cite{S10,line1,line2,line3})
\begin{equation}
\begin{split}
\partial_{t}D=SD+DS^{\top}+N
\label{eq:env}
\end{split}
\end{equation}
$N$ in Eq. (\ref{eq:env}) is a noise matrix  and  it will be a diagonal form, i.e., $\textbf{diag}(\kappa, \kappa, 0, \gamma_{1}(2\bar{n}_b+1), 0, \gamma_{2}(2\bar{n}_b+1))$, if the noise  correlation is defined by $\langle\hat{\zeta}_i(t)\hat{\zeta}_j(t)+\hat{\zeta}_j(t)\hat{\zeta}_i(t)\rangle/2= N_{ij}\delta(t-t')$. With the help of  Eq. (\ref{eq:env}),  above mentioned synchronization measurement $S'_{g}$ can be simply expressed as
\begin{equation}
\begin{split}
S'_g(t)=&\langle\delta q_{-}^2(t)+\delta p_{-}^2(t)\rangle^{-1}\\
       = &\{\dfrac{1}{2}[D_{33}(t)+D_{55}(t)-2D_{35}(t)]\\
       &+\dfrac{1}{2}[D_{44}(t)+D_{66}(t)-2D_{46}(t)]\}^{-1}
\label{eq:Scenv}
\end{split}
\end{equation}
and its evolution can be obtained by solving Eq. (\ref{eq:mean}) and Eq. (\ref{eq:env}) in order. At this point, all dynamic properties of our system, including synchronization and correlation, can be learned by means of the solutions of Eq. (\ref{eq:mean}), (\ref{eq:env}) and (\ref{eq:Scenv}). In the following subsections, we introduce two common forms of generalized synchronization: constant error synchronization (\ref{Constant error synchronization}) and time delay synchronization (\ref{Time delay synchronization}) to exhibit our ability about control synchronization, we will also prove how the controller is designed  to realize these synchronizations.

\subsection{Constant error synchronization}
\label{Constant error synchronization}
Constant error synchronization can be regarded as a translation in phase space between two systems. In Eq. (\ref{eq:classc}), if we let $h_{1}(x_1)=x_1+c_1$ and $h_{2}(x_2)=x_2+c_2$, then the ``classical'' synchronization criterion will be  
\begin{equation}
\begin{split}
\lim_{t\rightarrow\infty}\left|x_1(t)-x_2(t)\right|\rightarrow c_2-c_1=c_{-}
\end{split}
\label{eq:constant}
\end{equation}
where $c_{-}$ is the so-called constant error. In view of the controller influencing directly on $\partial_{t}p_{j}$, we firstly only consider the evolutions of the momentum operators, and further construct following Lyapunov function by using their expectation values:
\begin{equation}
\begin{split}
V_{p}(t)=(p_1(t)-p_2(t))^2
\label{eq:lyp}
\end{split}
\end{equation}
One can easily verify that $V_{p}\geqslant 0$ and $V_{p}=0$ is valid only when $p_1(t)-p_2(t)=0$. Substituting Eq. (\ref{eq:mean}) into Eq.(\ref{eq:lyp}), the time derivative of $V_{p}$ can be calculated handily if $C_1(t)=C_2(t)=C(t)$
\begin{equation}
\begin{split}
\dot{V}_{p}(t)=&2(\dot{p}_1(t)-\dot{p}_2(t))(p_1(t)-p_2(t))\\
              =&2\{[1+C(t)](\omega_{m2}q_{2}-\omega_{m1}q_{1})-\gamma_{1}p_{1}\\
               &+\gamma_{2}p_{2}+(g_{1}-g_{2})\vert A\vert^2\}(p_1-p_2)
\label{eq:dotlyp}
\end{split}
\end{equation}
we find that $\dot{V}_{p}(t)$ is always non--positive by setting 
\begin{equation}
\begin{split}
\dot{p}_1(t)-\dot{p}_2(t)=-k(p_1(t)-p_2(t))
\label{eq:ddotlyp}
\end{split}
\end{equation}
where $k$ is a positive real number. With this choice, $V_p$ simultaneously satisfies $V_{p}\geqslant 0$ and $\dot{V}_{p}=-2k(p_1(t)-p_2(t))^2\leqslant 0$. Under this condition, the system will gradually evolve to a stable state which corresponds to the origin of the Lyapunov function, i.e,
$p_1(t)=p_2(t)$ \cite{Con}. In order to satisfy the required form of Lyapunov function, the control field
can be obtained based on Eqs. (\ref{eq:dotlyp}) and (\ref{eq:ddotlyp})
\begin{equation}
\begin{split}
C(t)=\dfrac{(\gamma-k)[p_1-p_2]-(g_1-g_2)\vert A\vert^2}{\omega_{m2}q_{2}-\omega_{m1}q_{1}}-1
\label{eq:confield}
\end{split}
\end{equation}
here we have already set $\gamma_{1}=\gamma_{2}=\gamma$. It is needed to emphasize that all mechanical quantities without specifically being marked in this equation represent the expectation values at time $t$ (e.g., $p_1:=p_1(t)$). Otherwise, confusions may occur in following discussion about time delay synchronization.

We notice, however, the control field $C(t)$ in Eq. (\ref{eq:confield}) could be infinite when the tracks of $q_1$ and $q_2$ are adjacent. In particular, complete synchronization is not acceptable if $\omega_{m1}\simeq \omega_{m2}$. For avoiding this singularity, it is necessary to add a lower bound in the denominator of the control field. Therefore, the control field is modified in follow form:
\begin{equation}
C(t)=\left\{
\begin{aligned}
&\dfrac{(\gamma-k)[p_1-p_2]-(g_1-g_2)\vert A\vert^2}{\omega_{m2}q_{2}-\omega_{m1}q_{1}}-1 \\
&(when\,\,\vert\omega_{m2}q_{2}-\omega_{m1}q_1\vert>c_-)\\
&0 \\
&(when\,\,\vert\omega_{m2}q_{2}-\omega_{m1}q_{1}\vert\leqslant c_-)\\
\end{aligned}
\right.
\label{eq:confin}
\end{equation}
\begin{figure}[b]
\centering
\begin{minipage}{0.48\textwidth}
\centering
\includegraphics[width=3.3in]{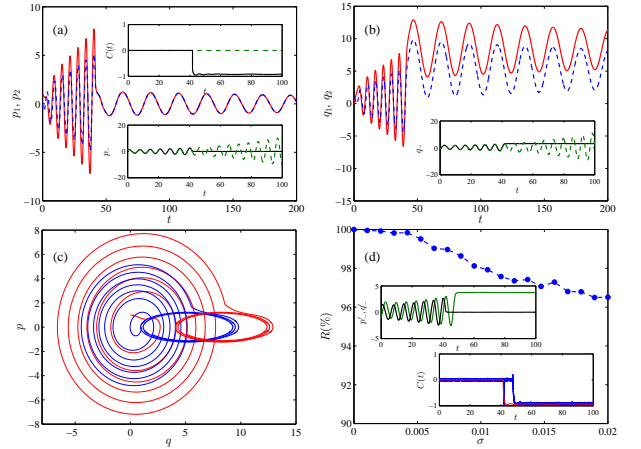}  
\end{minipage}
\caption{Evolutions of expectation values (blue dashed and red solid), control field and errors (green dashed and black solid respectively denote control field is imposed or removed), in which (a) and (b) respectively corresponds to momentum and position operators of each oscillator. Here we set $\Delta=1$ as a unit and other parameters are: $\omega_{m1}=1$, $\omega_{m2}=1.005$, $g_{1}=0.008$, $g_{2}=0.005$, $E=10$,  $\kappa=0.15$, $\gamma=0.005$ and $\bar{n}_b=0.05$. For the control field, the parameters are taken as  $k=2$ and $c_-=3$. (c): The limit cycles of two oscillators in phase space. (d): Robustness of our control system.
\label{fig:fig2}}
\end{figure}

The physical mechanism corresponding to Eq. (\ref{eq:confin}) can be interpreted as follows: Assuming the gap between two oscillators is small enough to satisfy $\vert\omega_{m2}q_{2}-\omega_{m1}q_{1}\vert\leqslant c_-$ at initial moment, then the control field will not work and the difference between oscillators under different Hamiltonians will increase; Once such difference crosses the boundary $\vert\omega_{m2}q_{2}-\omega_{m1}q_{1}\vert> c_-$, then non--zero control field will drag their orbits to close each other until a critical distance is reached, which will cause the invalid control field is resumed again.
With time going by, the error evolution will be controlled in a stable limit ellipse. Under particular $k$ and $c_-$, it can be regard as a fixed point if the major axis of this ellipsoid is small enough. In this case, two systems will finally realize such a synchronization: $p_{1}-p_{2}=0$ and $q_{1}-q_{2}=c_-$. Therefore, we make sure that $C(t)$ in Eq. (\ref{eq:confield}) is able to control the system achieving generalized synchronization. 

In Fig. \ref{fig:fig2} we provide simulation results of the two oscillators to verify the synchronization phenomenon under the control field. In Fig. \ref{fig:fig2}(a), one can directly see momentums of two oscillators will take on consistent evolution after $t=41.3$, which is exactly the same with the time point that control field is non-zero. Correspondingly, the momentum error will stabilize at zero instead of generally enlarging along with the control field. Fig. \ref{fig:fig2}(a) also quantitatively shows the control field is a slowly varying function of the time. Such a slowly varing control field can improve the stability of the system, simultaneously, it is more easily to be implemented by experiments. In Fig. \ref{fig:fig2}(b), we plot the positions of two oscillators and corresponding errors. It illustrates that, although two oscillators are not consistent in their positions, the error can still maintain a constant ($c_-$). Taken Fig. \ref{fig:fig2}(a) and Fig. \ref{fig:fig2}(b) together, we can determine constant error synchronization between two oscillators is achieved. In Fig. \ref{fig:fig2}(c), we show the ``tracks" of two oscillators in phase space. Two oscillators will evolve to their respective limit cycle and, as we predicted above, constant error synchronization corresponds to a translation between limit cycles in phase space. Fig. \ref{fig:fig2}(d) reports the robustness of our synchronization system. Here we assume each quantity in Eq. (\ref{eq:confin}) has been added a Gaussian noise whose standard deviation is $\sigma$, i.e., $o(t)=\mathcal{N}(o(t),\sigma)$ ($o\in\{q_{1,2},p_{1,2},A$) and when the final control field is
added on the system, it also has a noise ($C(t)=\mathcal{N}(C(t),\sigma)$). The accuracy of the synchronization scheme in this case is described by following auxiliary quantity
\begin{equation}
\begin{split}
R(\sigma)=1-\dfrac{[(q_{-}-q'_{-})^2+(p_{-}-p'_{-})^2]^{1/2}}{\sqrt{2}r}
\label{eq:lubang}
\end{split}
\end{equation}
where $p'_{-},q'_{-}$ refer to the errors in biased control field, and $r$ is the average radius of the limit cycle. One can find that $R(\sigma)$ will always remain above $96\%$ even the $\sigma =0.02$. Under those parameters, even if there are obvious fluctuations in the control field, however, the errors between two oscillators are still stable to approach $0, c_-'$. Therefore, we confirm our control is stable enough for some interferences.
\begin{figure}[]
\centering
\begin{minipage}{0.48\textwidth}
\centering
\includegraphics[width=3.3in]{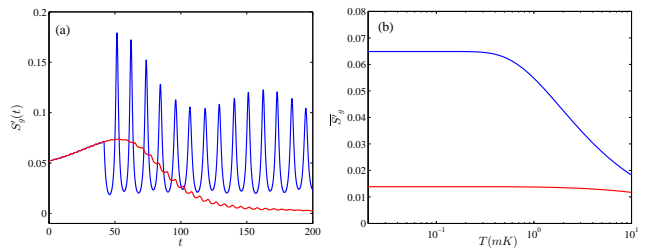}  
\end{minipage}
\caption{(a): Evolution of modified synchronization measurement. (b): Time-averaged synchronization measurement with varied bath temperature. Here blue of corresponds to control field imposed and red is the case control field disappears.
\label{fig:fig3}}
\end{figure}

Besides the expectation values, Mari's measurement is also calculated to prove that quantum fluctuation is  similarly squeezed by control field. Fig. \ref{fig:fig3}(a) illustrates an increasing $S'_g$ substitutes the trend to $0$, which significantly outperforms uncontrolled situation. Therefore, we recognize that the control field can indeed achieve quantum control rather than the synchronization of the classical level. In Fig. \ref{fig:fig3}(b), we also show how the bath temperature will influence synchronization phenomenon and it can be known that $S'_g(t)$ will keep almost unchanging if the bath temperature is limited within $1mK$($\bar{n}_b=0.28$ corresponding to a MHz phonon frequency), and it is still larger than that belonging to the uncontrolled system even though $T$ goes up to $10mK$($\bar{n}_b=6.14$). This range is quite broad compared to other correlation control schemes in optomechanical systems \cite{te1,te2}. 

\subsection{Time delay synchronization}
\label{Time delay synchronization}
Time delay synchronization can be regarded as a constant phase deviation between two systems, and their ``tracks" in phase space are overlapping like complete synchronization. In Eq. (\ref{eq:classc}), if we set $h_{1}(x_1)=x_1(t)$ and $h_{2}(x_2)=x_2(t-\tau)$, then the ``classical'' synchronization criterion will be  
\begin{equation}
\begin{split}
\lim_{t\rightarrow\infty}\left|x_1(t)-x_2(t-\tau)\right|\rightarrow 0
\end{split}
\label{eq:timedecay}
\end{equation}
Similarly with the above discussion, we define following Lyapunov function: 
\begin{equation}
\begin{split}
V_{p}(t)=(p_1(t)-p_2(t-\tau))^2
\label{eq:lyptd}
\end{split}
\end{equation}
and its derivative can also be expressed as $\dot{V}_{p}=2(\dot{p}_1-\dot{p}_2(t-\tau))(p_1-p_2(t-\tau))$, where
\begin{equation}
\begin{split}
\dot{p}_1-&\dot{p}_2(t-\tau)=-\omega_{m1}[1+C_1]q_{1}-\gamma_{1}p_{1}+g_{1}\vert A\vert^2\\
               &+\omega_{m2}q_{2}(t-\tau)+\gamma_{2}p_{2}(t-\tau)+g_{2}\vert A(t-\tau)\vert^2
\label{eq:dotlyptd}
\end{split}
\end{equation}

It needs to emphasize again that, in above expressions, all mechanical quantities without specifically being marked represent the expectation values at time $t$. Similarly, we let
\begin{equation}
\begin{split}
\dot{p}_1-&\dot{p}_2(t-\tau)=-k(p_1(t)-p_2(t-\tau))
\label{eq:ddotlyptd}
\end{split}
\end{equation}
to satisfy $V_{p}\geqslant 0$ and $\dot{V}_{p}=-2k(p_1-p_2(t-\tau))^2\leqslant 0$. Then the corresponding control field will become:

\begin{widetext}
\begin{equation}
\begin{split}
C_1(t)=\dfrac{(\gamma-k)[p_1-p_2(t-\tau)]-g_1\vert A\vert^2+g_2\vert A(t-\tau)\vert^2-\omega_{m2}q_{2}(t-\tau)}{-\omega_{m1}q_{1}}-1
\label{eq:confieldtdt}
\end{split}
\end{equation}
via setting $C_2(t)=0$ and $\gamma_{1}=\gamma_{2}=\gamma$ for simplicity.

Eq. (\ref{eq:confieldtdt}) is also of singular point at $q_1(t)=0$, therefore, an artificial boundary is necessary to avoid an infinite control field too. Unlike Eq. (\ref{eq:confin}), our purpose here is to make two systems achieve complete synchronization after eliminating the time delay. Therefore, this limitation is on the whole control field instead of the denominator. Then the control field should be
\begin{equation}
C_1(t)=\left\{
\begin{aligned}
&\dfrac{(\gamma-k)[p_1-p_2(t-\tau)]-g_1\vert A\vert^2+g_2\vert A(t-\tau)\vert^2-\omega_{m2}q_{2}(t-\tau)}{-\omega_{m1}q_{1}}-1
\,\,\,\,\,\,\,\,&(-C_M\leqslant C_1\leqslant C_M)\\
&C_M 
&(C_1 > C_M)\\
&-C_M 
&(C_1 < -C_M)\\
\end{aligned}
\right.
\label{eq:confintdt}
\end{equation}
\end{widetext}

\begin{figure}[b]
\centering
\begin{minipage}{0.48\textwidth}
\centering
\includegraphics[width=3.3in]{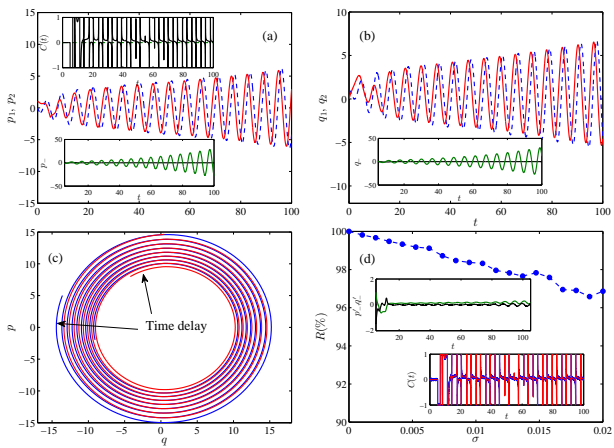}  
\end{minipage}
\caption{Evolutions of expectation values (blue dashed and red solid), control field and errors (green dashed and black solid respectively denote control field is imposed or removed), in which (a) and (b) respectively corresponds to momentum and position operators of each oscillator. (c): The limit cycles of two oscillators in phase space. (d): Robustness of control system. Here we set $\tau=5$, $C_M=1$ and other parameters are the same with Fig. \ref{fig:fig2}.
\label{fig:fig4}}
\end{figure}

In Fig. \ref{fig:fig4}(a) and (b), 
we show that the evolution of one oscillator seems to be a time translation of the other oscillator, and the errors tend to zero like complete synchronization after eliminating the time delay. It also exhibits a quickly varying control field, which is different with the performance in constant error synchronization. Generally speaking, quickly varying control field
can make the system achieve synchronization faster. Fig. \ref{fig:fig4}(a) and (b) show two oscillators will achieve synchronization in a short period of time $t<10$, which shortens four times time relative to  Fig. \ref{fig:fig2} (a) and (b). Furthermore, Fig. \ref{fig:fig4}(d) illustrates synchronization accuracy is $96\%$, which means robustness is also remained in a high level under the fast oscillating control field. We also plot the limit cycles of two oscillators in Fig. \ref{fig:fig4}(c). It can be seen that two limit cycles are almost coincident in most of time except the short time intervals at origin and destination points in which both limit circles take on inconsistent evolutions because of the time delay.
\begin{figure}[b]
\centering
\begin{minipage}{0.48\textwidth}
\centering
\includegraphics[width=3.3in]{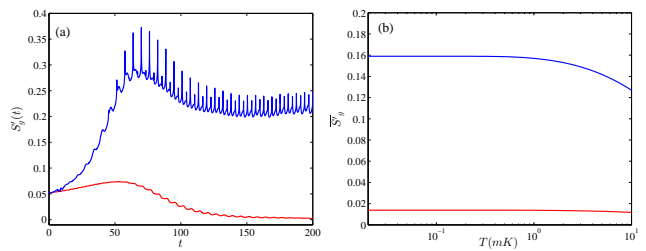}  
\end{minipage}
\caption{(a): Evolution of modified synchronization measurement. (b): Time-averaged synchronization measurement with varied bath temperature. Here blue corresponds to control field imposed and red is the case control field disappears.
\label{fig:fig5}}
\end{figure}

We also consider the evolution of quantum fluctuation. Fig. \ref{fig:fig5}(a) shows that the fluctuations are further squeezed by quickly varying control field, and $S'_g(t)$ exhibits a higher value than that in constant error synchronization. Fig. \ref{fig:fig5}(b) also illustrates that destruction on the synchronization effect causing by environment is also weakened and $\bar{S}'_g(t)$ will still remain high even at $T=10mK$. From this perspective, we believe that quickly varying control field is a appropriate form of synchronization control too.
 
\section{correlation in Generalized synchronization} 
\label{correlation in Generalized synchronization} 
The correlation between synchronized quantum systems is a important research object in QIP. Intuitively, two different systems can achieve consistency in some extent, meaning that there inevitably exists a certain correlation between those systems. To verify this, quantum mutual information which is a measurement of total correlation has been proved to be homology with synchronization measurement in VdP oscillators \cite{S2}. However, it is very difficult to identify the type of this correlation. Especially, whether quantum entanglement exists in the CV synchronization is controversial \cite{S1,S2,S6,S9}. Therefore, we pay more attention to the properties of entanglement when we consider the quantum correlation in our model. 

The mean--field approximation used above can make it more convenient to analyze classical correlation and quantum entanglement. The classical correlation can be verified via calculating the largest Lyapunov exponent of the errors \cite{S10,line2}, i.e., $L_{y}^{max}=\max\{L_y(p_{g-}),L_y(q_{g-})\}$, where 
\begin{equation}
\begin{split}
L_y(o)=\lim_{t\rightarrow\infty}\dfrac{1}{t}\ln\left|\dfrac{\delta o(t)}{\delta o(0)}\right|,\,\,\,\,\,\, (o\in\{p_{g-},q_{g-}\})
\label{eq:lizhishu}
\end{split}
\end{equation}
Correspondingly, CV quantum entanglement is measured by Gaussian Negativity $E_n=\max\{0,-\log_{2}\nu_{-}\}$ \cite{Na1,Na2,LE3}, here
\begin{equation}
\begin{split}
\nu_{-}=\dfrac{\sqrt{\Delta(\Gamma)-\sqrt{\Delta(\Gamma)^2-4\det\Gamma}}}{2},
\label{eq:enenen}
\end{split}
\end{equation}
$\Delta(\Gamma)=\det A+\det B-2\det C$ and 
\begin{equation}
\Gamma=\begin{pmatrix}
 D_{33}& D_{34}& D_{35}&  D_{36}\\ 
 D_{43}& D_{44}& D_{45}&  D_{46}\\ 
 D_{53}& D_{54}& D_{55}&  D_{56}\\ 
 D_{63}& D_{64}& D_{65}&  D_{66}
 \end{pmatrix}
 =\left( \begin{array}{ll} 
 A&  C\\ 
 C^{\top}&  B
 \end{array}
\right).
\label{eq:enenenmatrix}
\end{equation}
\begin{figure}[]
\centering
\begin{minipage}{0.48\textwidth}
\centering
\includegraphics[width=3.3in]{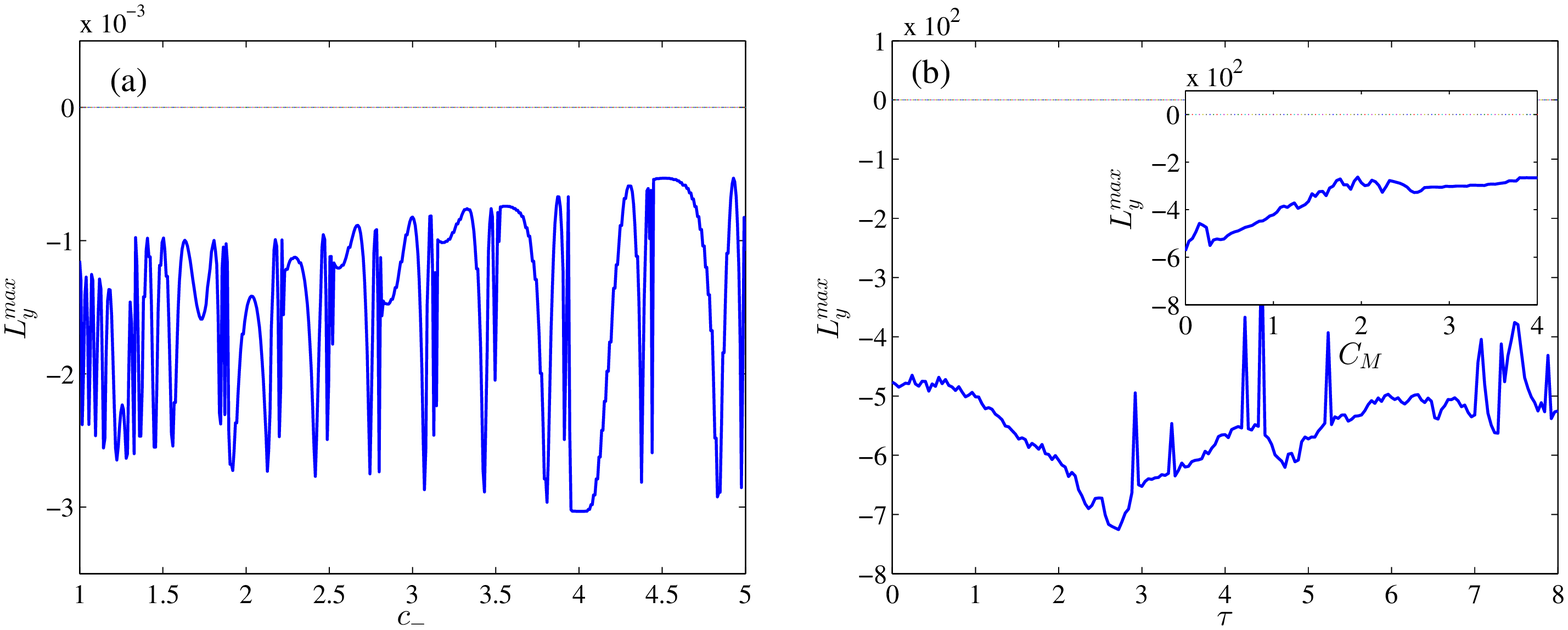}  
\end{minipage}
\caption{The largest Lyapunov exponents of the errors with varied $c_-$ (a), $\tau$ and $C_M$ (b) which corresponds to constant error synchronization and time delay synchronization, respectively.
\label{fig:fig6}}
\end{figure}

In Fig. \ref{fig:fig6}, we plot the largest Lyapunov exponent under different characteristic parameters ($c_-$, $\tau$ and $C_M$). The results show that negative Lyapunov exponent is always present under different control fields. This performance is superior to our conclusion in Ref. \cite{S10}, in which we found that the positive-negative of Lyapunov exponent corresponding to coupling synchronization depends sensitively on varing coupling intensity between two opomechanical systems. This means, compared to the direct coupling between two systems, designed Lyapunov function can more effectively help the system to establish correlation in the expect value level.

\begin{figure}[]
\centering
\begin{minipage}{0.48\textwidth}
\centering
\includegraphics[width=3.3in]{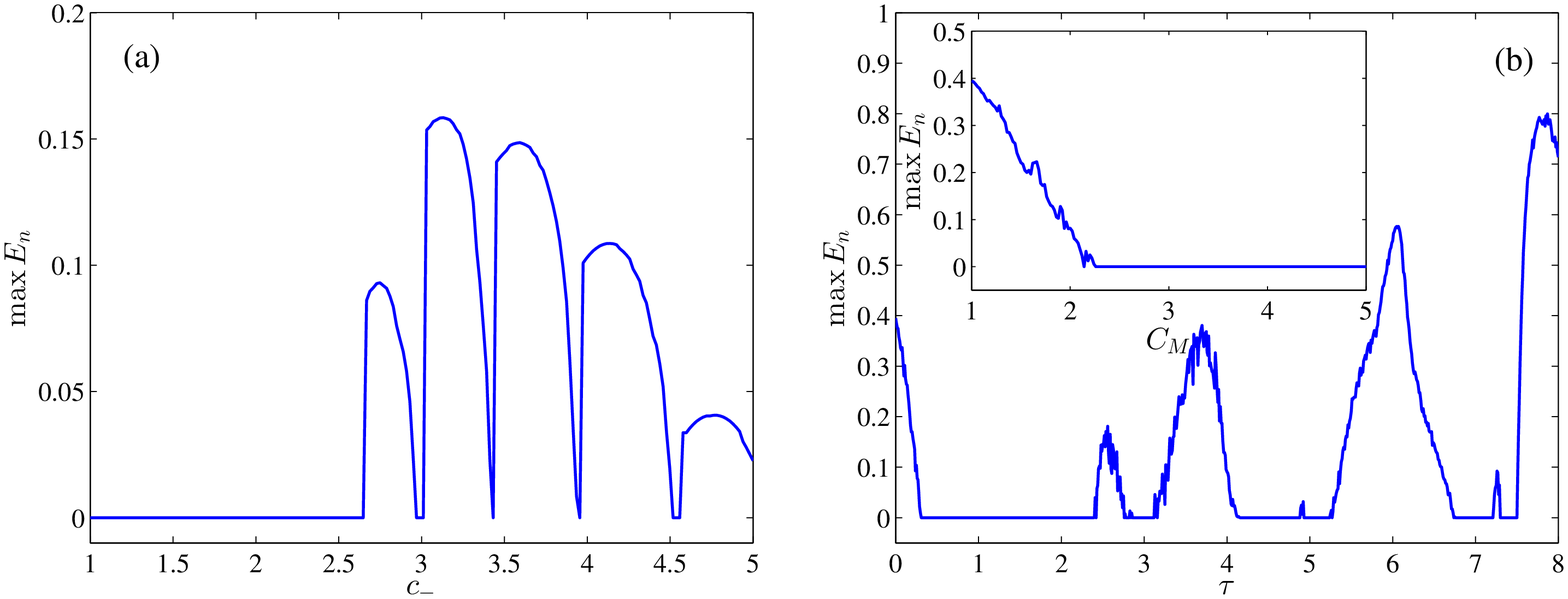}  
\end{minipage}
\caption{ The maximum Negativity of the errors with varied $c_-$ (a), $\tau$ and $C_M$ (b) which corresponds to constant error synchronization and time delay synchronization, respectively.
\label{fig:fig7}}
\end{figure}

In Fig. \ref{fig:fig7}, we plot the results of maximum Negativity under different characteristic parameters. It should be noted that generalized synchronization is actually a necessary condition for complete synchronization and characteristic parameters can be regarded as a description about the gap between generalized synchronization and complete synchronization. Consequently, one can compare complete synchronization with generalized synchronization via setting $c_-=0$, $\tau=0$ and $C_M\rightarrow\infty$. Fig. \ref{fig:fig7}(a) shows that quantum entanglement does not exist until $c_-=2.65$. Then   
it takes on a rising trend and its the maximum value is at $c_-=3.13$. Subsequently, Negativity will decline and tend to zero again. The evolution of maximum Negativity can be understood as follow: when $c_-$ is small enough, the oscillators are equivalent to achieving complete synchronization and they are always separable in this case. This opinion is the same with Mari's conclusion, i.e., the evolution in superposition tracks will hinder the generation of CV entanglement. With gradually going away of
two tracks, entanglement is allowed to exist because the oscillators are no longer complete synchronization. Finally, if the distance of two tracks increases constantly, quantum correlation between two oscillators is remarkably weakened and entanglement will disappeare again when $c_-$ is large. In Fig. \ref{fig:fig7}(b), similar conclusions are also obtained in time delay synchronization. Therefore, we think that Gaussian entanglement can coexist with generalized synchronization, although it may be prohibited by complete synchronization.

\section{discussion and Results}
\label{Results and discussion}  
Here we give a brief discussion about the parameters of our optpmechanical system and the realization scheme of the control field. The parameters selected in the simulation are similar
with Ref. \cite{S1,S9,Exx1,Exx2,line1}. However, in order to highlight the roles of the coupling and the controller, we appropriately reduce the value of the driving intensity \cite{S10}. Beyond that, the deviation in potential like Eq. (\ref{eq:contorl}) has been investigated by theoretical researches \cite{LE2}, and recent works report that the deviation can be achieved by using charged mechanical resonators \cite{MR1,MR2,MR3,MR4}. For example, \citeauthor{MR2} found that the effective frequency $\omega_{eff}^2=\omega_{m}^2[(1+C(t))]$ can be controlled by a time-dependent bias gate voltage $U(t)=U_0f(t)$. And the relationship between dimensionless factor $f(t)$ and $C(t)$ is $C(t)=\eta f(t)$, where 
\begin{equation}
\eta=\dfrac{C_0U_0Q_{MR}}{\pi\varepsilon_0m\omega_m^2d^3}
\label{eq:contorlexex}
\end{equation}
is obtained in Ref. \cite{MR2}. Therefore, we are sure the control terms in Eq. (\ref{eq:confin}) and (\ref{eq:confintdt}) cab be achieved easily in a specific experiment.

In summary, we have extend Mari's theories about quantum complete synchronization and phase synchronization to a more general situation that we defined as quantum generalized synchronization in this paper. The corresponding control methods, criteria and measurements are also proposed quantificationally based on Lyapunov function, Lyapunov exponent and modified Mari's measurement. This generalized synchronization can be regard as a prerequisite of traditional quantum synchronization, and it can establish more flexible relations between two controlled systems. To verify this, we have shown that some important properties in our model, such as entanglement in synchronization, will be consistent with previous works if the generalized synchronization tends to complete synchronization. So, designers can complete different synchronizations according to their requirements based on our theory. For making our theory more intuitive, we have considered two common generalized synchronization, that is, so--called  constant error synchronization and time delay synchronization in an optomechanical system. With the help of control fields designed by Lyapunov function, we have proved two oscillators can satisfy the requirements of various synchronizations. We believe that our work can bring certain application values in quantum information transmission, quantum control, and quantum logical processing.  
\section*{ACKNOWLEDGEMENT} 
All authors thank Jiong Cheng, Wenzhao Zhang and Yang Zhang for the useful discussion. This research was supported by the National Natural Science Foundation of China (Grant No 11175033) and the Fundamental Research Funds for the Central Universities (DUT13LK05).

\appendix
\section{Parameters in quantum Langevin equations}
\label{Parameters in quantum Langevin equations}
The concrete form of coefficient matrix $S$ in Eq. (\ref{eq:env}) is 
\begin{widetext}
\begin{equation}
S=
\begin{pmatrix}
 -\kappa&  -(\Delta+g_1q_1+g_2q_2)& -\sqrt{2}g_1\text{Im}(A)&  0& -\sqrt{2}g_2\text{Im}(A)&  0\\ 
 \Delta+g_1q_1+g_2q_2&  -\kappa& \sqrt{2}g_1\text{Re}(A)&  0& \sqrt{2}g_2\text{Re}(A)&  0\\ 
 0&  0& 0&  \omega_{m1}& 0&  0\\ 
 \sqrt{2}g_1\text{Re}(A)&  \sqrt{2}g_1\text{Im}(A)& -\omega_{m1}[1+c_{1}(t)]&  -\gamma_{1}& 0&  0\\ 
 0&  0& 0&  0& 0&  \omega_{m2}\\ 
\sqrt{2}g_2\text{Re}(A)&  \sqrt{2}g_2\text{Im}(A)& 0& 0& -\omega_{m2}[1+c_{2}(t)]& -\gamma_{2}
 \end{pmatrix}
\label{eq:enenenmatrix}
\end{equation}
\end{widetext}

\begin{thebibliography}{35}
\expandafter\ifx\csname
natexlab\endcsname\relax\def\natexlab#1{#1}\fi
\expandafter\ifx\csname bibnamefont\endcsname\relax
  \def\bibnamefont#1{#1}\fi
\expandafter\ifx\csname bibfnamefont\endcsname\relax
  \def\bibfnamefont#1{#1}\fi
\expandafter\ifx\csname citenamefont\endcsname\relax
  \def\citenamefont#1{#1}\fi
\expandafter\ifx\csname url\endcsname\relax
  \def\url#1{\texttt{#1}}\fi
\expandafter\ifx\csname
urlprefix\endcsname\relax\def\urlprefix{URL }\fi
\providecommand{\bibinfo}[2]{#2}
\providecommand{\eprint}[2][]{\url{#2}}

\bibitem[{\citenamefont{Mari et~al.}(2014)\citenamefont{Mari}}]{S1}
\bibinfo{author}{\bibfnamefont{A.} \bibnamefont{Mari}},
\bibinfo{author}{\bibfnamefont{A.} \bibnamefont{Farace}},
\bibinfo{author}{\bibfnamefont{N.} \bibnamefont{Didier}},
\bibinfo{author}{\bibfnamefont{V.} \bibnamefont{Giovannetti}}
\bibnamefont{and} \bibinfo{author}{\bibfnamefont{R.}~\bibnamefont{Fazio}},
\bibinfo{journal}{Phys. Rev. Lett.}
\textbf{\bibinfo{volume}{111}}, \bibinfo{pages}{103605-5}
(\bibinfo{year}{2013}).

\bibitem[{\citenamefont{Ameri et~al.}(2014)\citenamefont{Ameri, V. and Eghbali-Arani, M. and Mari, A. and Farace, A. and Kheirandish, F. and Giovannetti, V. and Fazio, R.}}]{S2}
\bibinfo{author}{\bibfnamefont{V.} \bibnamefont{Ameri}},
\bibinfo{author}{\bibfnamefont{M.} \bibnamefont{Eghbali-Arani}},
\bibinfo{author}{\bibfnamefont{A.} \bibnamefont{Mari}},
\bibinfo{author}{\bibfnamefont{A.} \bibnamefont{Farace}},
\bibinfo{author}{\bibfnamefont{F.} \bibnamefont{Kheirandish}},
\bibinfo{author}{\bibfnamefont{V.} \bibnamefont{Giovannetti}}
\bibnamefont{and} \bibinfo{author}{\bibfnamefont{R.}~\bibnamefont{Fazio}},
\bibinfo{journal}{Phys. Rev. A}
\textbf{\bibinfo{volume}{91}}, \bibinfo{pages}{012301-6}
(\bibinfo{year}{2015}).

\bibitem[{\citenamefont{Xu et~al.}(2014)\citenamefont{Xu}}]{S3}
\bibinfo{author}{\bibfnamefont{M.~H.} \bibnamefont{Xu}},
\bibinfo{author}{\bibfnamefont{D.~A.} \bibnamefont{Tieri}},
\bibinfo{author}{\bibfnamefont{E.~C.} \bibnamefont{Fine}},
\bibinfo{author}{\bibfnamefont{J.~K.} \bibnamefont{Thompson}}
\bibnamefont{and} \bibinfo{author}{\bibfnamefont{M.~J.}~\bibnamefont{Holland}},
\bibinfo{journal}{Phys. Rev. Lett.}
\textbf{\bibinfo{volume}{113}}, \bibinfo{pages}{154101-5}
(\bibinfo{year}{2014}).

\bibitem[{\citenamefont{Xu et~al.}(2015)\citenamefont{Xu}}]{S4}
\bibinfo{author}{\bibfnamefont{M.~H.} \bibnamefont{Xu}}
\bibnamefont{and} \bibinfo{author}{\bibfnamefont{M.~J.}~\bibnamefont{Holland}},
\bibinfo{journal}{Phys. Rev. Lett.}
\textbf{\bibinfo{volume}{114}}, \bibinfo{pages}{103601-5}
(\bibinfo{year}{2015}).

\bibitem[{\citenamefont{Lee et~al.}(2014)\citenamefont{Lee, Tony E. and Sadeghpour, H. R}}]{S5}
\bibinfo{author}{\bibfnamefont{T.~E.} \bibnamefont{Lee}}
\bibnamefont{and} \bibinfo{author}{\bibfnamefont{H.~R.}~\bibnamefont{Sadeghpour}},
\bibinfo{journal}{Phys. Rev. Lett.}
\textbf{\bibinfo{volume}{111}}, \bibinfo{pages}{234101-5}
(\bibinfo{year}{2013}).

\bibitem[{\citenamefont{Lee et~al.}(2014)\citenamefont{Lee, Tony E. and Chan, Ching-Kit and Wang, Shenshen}}]{S6}
\bibinfo{author}{\bibfnamefont{T.~E.} \bibnamefont{Lee}},
\bibinfo{author}{\bibfnamefont{C.~K.} \bibnamefont{Chan}}
\bibnamefont{and} \bibinfo{author}{\bibfnamefont{S.~S.}~\bibnamefont{Wang}},
\bibinfo{journal}{Phys. Rev. E}
\textbf{\bibinfo{volume}{89}}, \bibinfo{pages}{022913-10}
(\bibinfo{year}{2014}).

\bibitem[{\citenamefont{Walter et~al.}(2014)\citenamefont{Walter, Stefan and Nunnenkamp, Andreas and Bruder, Christoph}}]{S11}
\bibinfo{author}{\bibfnamefont{S.} \bibnamefont{Walter}},
\bibinfo{author}{\bibfnamefont{A.} \bibnamefont{Nunnenkamp}}
\bibnamefont{and} \bibinfo{author}{\bibfnamefont{C.}~\bibnamefont{Bruder}},
\bibinfo{journal}{Phys. Rev. Lett.}
\textbf{\bibinfo{volume}{112}}, \bibinfo{pages}{094102-5}
(\bibinfo{year}{2014}).

\bibitem[{\citenamefont{Samoylova et~al.}(2014)\citenamefont{Marina Samoylova, Nicola Piovella, Gordon R.M. Robb, Romain Bachelard, Philippe W. Courteille}}]{S7}
\bibinfo{author}{\bibfnamefont{M.} \bibnamefont{Samoylova}},
\bibinfo{author}{\bibfnamefont{N.} \bibnamefont{Piovella}},
\bibinfo{author}{\bibfnamefont{G.~R.~M.} \bibnamefont{Robb}},
\bibinfo{author}{\bibfnamefont{R.} \bibnamefont{Bachelard}}
\bibnamefont{and} \bibinfo{author}{\bibfnamefont{P.~W.}~\bibnamefont{Courteille}},
\bibinfo{journal}{arXiv: 1503.05616v1}
(\bibinfo{year}{2015}).

\bibitem[{\citenamefont{Samoylova et~al.}(2014)\citenamefont{Marina Samoylova, Nicola Piovella, Gordon R.M. Robb, Romain Bachelard, Philippe W. Courteille}}]{S8}
\bibinfo{author}{\bibfnamefont{Y.} \bibnamefont{Gul}},
\bibinfo{journal}{arXiv:1412.8497v1}
(\bibinfo{year}{2015}).

\bibitem[{\citenamefont{Ying et~al.}(2014)\citenamefont{Lei Ying, Ying-Cheng Lai,and Celso Grebogi}}]{S9}
\bibinfo{author}{\bibfnamefont{L.} \bibnamefont{Ying}},
\bibinfo{author}{\bibfnamefont{Y.~C.} \bibnamefont{Lai}}
\bibnamefont{and} \bibinfo{author}{\bibfnamefont{C.}~\bibnamefont{Grebogi}},
\bibinfo{journal}{Phys. Rev. A}
\textbf{\bibinfo{volume}{90}}, \bibinfo{pages}{053810-6}
(\bibinfo{year}{2014}).

\bibitem[{\citenamefont{Li et~al.}(2015)\citenamefont{Li, W.L. and Li, C. and Song .H.S}}]{S10}
\bibinfo{author}{\bibfnamefont{W.~L.} \bibnamefont{Li}},
\bibinfo{author}{\bibfnamefont{C.} \bibnamefont{Li}},
\bibnamefont{and} \bibinfo{author}{\bibfnamefont{H.~S.}~\bibnamefont{Song}},
\bibinfo{journal}{J. Phys. B: At. Mol. Opt. Phys.}
\textbf{\bibinfo{volume}{48}}, \bibinfo{pages}{035503-8}
(\bibinfo{year}{2015}).

\bibitem[{\citenamefont{Choi et~al.}(2014)\citenamefont{Sun-Ho Choi and Seung-Yeal Ha}}]{S12}
\bibinfo{author}{\bibfnamefont{S.~H.} \bibnamefont{Choi}}
\bibnamefont{and} \bibinfo{author}{\bibfnamefont{S.~Y.}~\bibnamefont{Ha}},
\bibinfo{journal}{J. Phys. A: Math. Theor.}
\textbf{\bibinfo{volume}{47}}, \bibinfo{pages}{355104-16}
(\bibinfo{year}{2014}).

\bibitem[{\citenamefont{Choi et~al.}(2014)\citenamefont{Yong Xu, Hua Wang, Yongge Li, Bin Pei}}]{GSS1}
\bibinfo{author}{\bibfnamefont{Y.} \bibnamefont{Xu}},
\bibinfo{author}{\bibfnamefont{H.} \bibnamefont{Wang}},
\bibinfo{author}{\bibfnamefont{Y.} \bibnamefont{Li}}
\bibnamefont{and} \bibinfo{author}{\bibfnamefont{B.}~\bibnamefont{Pei}},
\bibinfo{journal}{Commun. Nonlinear Sci. Numer. Simulat.}
\textbf{\bibinfo{volume}{19}}, \bibinfo{pages}{3735–-3744}
(\bibinfo{year}{2014}).

\bibitem[{\citenamefont{Choi et~al.}(2014)\citenamefont{Yong Xu, Hua Wang, Yongge Li, Bin Pei}}]{GSS2}
\bibinfo{author}{\bibfnamefont{X.} \bibnamefont{Wu}},
\bibinfo{author}{\bibfnamefont{H.} \bibnamefont{Wang}}
\bibnamefont{and} \bibinfo{author}{\bibfnamefont{H.}~\bibnamefont{Lu}},
\bibinfo{journal}{Nonlinear Analysis: Real World Applications}
\textbf{\bibinfo{volume}{13}}, \bibinfo{pages}{1441–-1450}
(\bibinfo{year}{2012}).

\bibitem[{\citenamefont{Lü et~al.}(2014)\citenamefont{Ling lu, Chengren Li, Liansong Chen, Linling Wei}}]{GSS3}
\bibinfo{author}{\bibfnamefont{L.} \bibnamefont{L\"u}},
\bibinfo{author}{\bibfnamefont{C.} \bibnamefont{Li}},
\bibinfo{author}{\bibfnamefont{L.} \bibnamefont{Chen}}
\bibnamefont{and} \bibinfo{author}{\bibfnamefont{L.}~\bibnamefont{Wei}},
\bibinfo{journal}{Commun. Nonlinear Sci. Numer. Simulat.}
\textbf{\bibinfo{volume}{19}}, \bibinfo{pages}{2843–-2849}
(\bibinfo{year}{2014}).

\bibitem[{\citenamefont{Choi et~al.}(2014)\citenamefont{Sun-Ho Choi and Seung-Yeal Ha}}]{Con}
\bibinfo{author}{\bibfnamefont{G.} \bibnamefont{Nicolis}}
\bibnamefont{and} \bibinfo{author}{\bibfnamefont{I.}~\bibnamefont{Prigogine}},
\bibinfo{journal}{\textit{self-organization in nonequilibrium systems}},
(\bibinfo{pages}{Wiley},
\bibinfo{pages}{New York},
\bibinfo{year}{1977}).

\bibitem[{\citenamefont{Choi et~al.}(2014)\citenamefont{S. C. Hou, M. A. Khan, and X. X. Yi Daoyi Dong and Ian R. Petersen}}]{QC1}
\bibinfo{author}{\bibfnamefont{X.~X.} \bibnamefont{Yi}},
\bibinfo{author}{\bibfnamefont{X.~L.} \bibnamefont{Huang}},
\bibinfo{author}{\bibfnamefont{C.~F.} \bibnamefont{Wu}}
\bibnamefont{and} \bibinfo{author}{\bibfnamefont{C.~H.}~\bibnamefont{Oh}},
\bibinfo{journal}{Phys. Rev. A}
\textbf{\bibinfo{volume}{80}}, \bibinfo{pages}{052316-5}
(\bibinfo{year}{2009}).

\bibitem[{\citenamefont{Choi et~al.}(2014)\citenamefont{S. C. Hou, M. A. Khan, and X. X. Yi Daoyi Dong and Ian R. Petersen}}]{QC2}
\bibinfo{author}{\bibfnamefont{S.~C.} \bibnamefont{Hou}},
\bibinfo{author}{\bibfnamefont{M.~A.} \bibnamefont{Khan}},
\bibinfo{author}{\bibfnamefont{X.~X.} \bibnamefont{Yi}},
\bibinfo{author}{\bibfnamefont{D.~Y.} \bibnamefont{Dong}}
\bibnamefont{and} \bibinfo{author}{\bibfnamefont{I.~R.}~\bibnamefont{Petersen}},
\bibinfo{journal}{Phys. Rev. A}
\textbf{\bibinfo{volume}{86}}, \bibinfo{pages}{022321-8}
(\bibinfo{year}{2012}).

\bibitem[{\citenamefont{Choi et~al.}(2014)\citenamefont{S. C. Hou, M. A. Khan, and X. X. Yi Daoyi Dong and Ian R. Petersen}}]{H}
\bibinfo{author}{\bibfnamefont{M.} \bibnamefont{Aspelmeyer}},
\bibinfo{author}{\bibfnamefont{T.~J.} \bibnamefont{Kippenberg}}
\bibnamefont{and} \bibinfo{author}{\bibfnamefont{F.}~\bibnamefont{Marquardt}},
\bibinfo{journal}{Rev. Mod. Phys.}
\textbf{\bibinfo{volume}{86}}, \bibinfo{pages}{1391--1452}
(\bibinfo{year}{2014}).

\bibitem[{\citenamefont{Farace et~al.}(2014)\citenamefont{Alessandro Farace and Vittorio Giovannetti}}]{LE2}
\bibinfo{author}{\bibfnamefont{A.} \bibnamefont{Farace}}
\bibnamefont{and} \bibinfo{author}{\bibfnamefont{V.}~\bibnamefont{Giovannetti}},
\bibinfo{journal}{Phys. Rev. A}
\textbf{\bibinfo{volume}{86}}, \bibinfo{pages}{013820-12}
(\bibinfo{year}{2012}).

\bibitem[{\citenamefont{Choi et~al.}(2014)\citenamefont{S. C. Hou, M. A. Khan, and X. X. Yi Daoyi Dong and Ian R. Petersen}}]{LE1}
\bibinfo{author}{\bibfnamefont{C.} \bibnamefont{Genes}},
\bibinfo{author}{\bibfnamefont{A.} \bibnamefont{Mari}},
\bibinfo{author}{\bibfnamefont{D.} \bibnamefont{Vitalii}}
\bibnamefont{and} \bibinfo{author}{\bibfnamefont{S.}~\bibnamefont{Tombesi}},
\bibinfo{journal}{Adv. At. Mol. Opt. Phys.}
\textbf{\bibinfo{volume}{57}}, \bibinfo{pages}{33-86}
(\bibinfo{year}{2009}).

\bibitem[{\citenamefont{Choi et~al.}(2014)\citenamefont{S. C. Hou, M. A. Khan, and X. X. Yi Daoyi Dong and Ian R. Petersen}}]{LE3}
\bibinfo{author}{\bibfnamefont{Y.~D.} \bibnamefont{Wang}}
\bibnamefont{and} \bibinfo{author}{\bibfnamefont{A.~A.}~\bibnamefont{Clerk}},
\bibinfo{journal}{Phys. Rev. Lett.}
\textbf{\bibinfo{volume}{110}}, \bibinfo{pages}{253601-5}
(\bibinfo{year}{2013}).

\bibitem[{\citenamefont{Choi et~al.}(2014)\citenamefont{X. W. Xu Y,Li}}]{Noise0}
\bibinfo{author}{\bibfnamefont{C.~K.} \bibnamefont{Law}}
\bibinfo{journal}{Phys. Rev. A}
\textbf{\bibinfo{volume}{51}}, \bibinfo{pages}{2537}
(\bibinfo{year}{1995}).

\bibitem[{\citenamefont{Choi et~al.}(2014)\citenamefont{Sun-Ho Choi and Seung-Yeal Ha}}]{Noise1}
\bibinfo{author}{\bibfnamefont{C.~W.} \bibnamefont{Gardiner}}
\bibnamefont{and} \bibinfo{author}{\bibfnamefont{P.}~\bibnamefont{Zoller}},
\bibinfo{journal}{\textit{Quantum Noise}},
(\bibinfo{pages}{Springer, Berlin},
\bibinfo{year}{2000}).

\bibitem[{\citenamefont{Choi et~al.}(2014)\citenamefont{Yong-Chun Liu Yu-Feng Shen Qihuang Gong and Yun-Feng Xiao}}]{Noise2}
\bibinfo{author}{\bibfnamefont{Y.~C.} \bibnamefont{Liu}},
\bibinfo{author}{\bibfnamefont{Y.~F.} \bibnamefont{Shen}},
\bibinfo{author}{\bibfnamefont{Q.~H.} \bibnamefont{Gong}}
\bibnamefont{and} \bibinfo{author}{\bibfnamefont{Y.~F.}~\bibnamefont{Xiao}},
\bibinfo{journal}{Phys. Rev. A}
\textbf{\bibinfo{volume}{89}}, \bibinfo{pages}{053821-8}
(\bibinfo{year}{2014}).

\bibitem[{\citenamefont{Choi et~al.}(2014)\citenamefont{X. W. Xu Y,Li}}]{Noise3}
\bibinfo{author}{\bibfnamefont{X.~W.} \bibnamefont{Xu}}
\bibnamefont{and} \bibinfo{author}{\bibfnamefont{Y.}~\bibnamefont{Li}},
\bibinfo{journal}{Phys. Rev. A}
\textbf{\bibinfo{volume}{91}}, \bibinfo{pages}{053852-8}
(\bibinfo{year}{2015}).

\bibitem[{\citenamefont{Choi et~al.}(2014)\citenamefont{A. Mari and J. Eisert}}]{line1}
\bibinfo{author}{\bibfnamefont{A.} \bibnamefont{Mari}}
\bibnamefont{and} \bibinfo{author}{\bibfnamefont{J.}~\bibnamefont{Eisert}},
\bibinfo{journal}{Phys. Rev. Lett.}
\textbf{\bibinfo{volume}{103}}, \bibinfo{pages}{213603-4}
(\bibinfo{year}{2009}).

\bibitem[{\citenamefont{Wang et~al.}(2014)\citenamefont{G. L. Wang, L. Huang, Y. C. Lai, and C. Grebogi}}]{line2}
\bibinfo{author}{\bibfnamefont{G.~L.} \bibnamefont{Wang}},
\bibinfo{author}{\bibfnamefont{L.} \bibnamefont{Huang}},
\bibinfo{author}{\bibfnamefont{Y.~C.} \bibnamefont{Lai}}
\bibnamefont{and} \bibinfo{author}{\bibfnamefont{C.}~\bibnamefont{Grebogi}},
\bibinfo{journal}{Phys. Rev. Lett.}
\textbf{\bibinfo{volume}{112}}, \bibinfo{pages}{110406-5}
(\bibinfo{year}{2014}).

\bibitem[{\citenamefont{Wang et~al.}(2014)\citenamefont{G. L. Wang, L. Huang, Y. C. Lai, and C. Grebogi}}]{line3}
\bibinfo{author}{\bibfnamefont{J.} \bibnamefont{Larson}}
\bibnamefont{and} \bibinfo{author}{\bibfnamefont{M.}~\bibnamefont{Horsdal}},
\bibinfo{journal}{Phys. Rev. A}
\textbf{\bibinfo{volume}{84}}, \bibinfo{pages}{021804(R)-4}
(\bibinfo{year}{2011}).

\bibitem[{\citenamefont{Wang et~al.}(2014)\citenamefont{G. L. Wang, L. Huang, Y. C. Lai, and C. Grebogi}}]{te1}
\bibinfo{author}{\bibfnamefont{J.~Q.} \bibnamefont{Liao}}
\bibnamefont{and} \bibinfo{author}{\bibfnamefont{F.}~\bibnamefont{Nori}},
\bibinfo{journal}{Phys. Rev. A}
\textbf{\bibinfo{volume}{88}}, \bibinfo{pages}{023853}
(\bibinfo{year}{2013}).

\bibitem[{\citenamefont{Wang et~al.}(2014)\citenamefont{G. L. Wang, L. Huang, Y. C. Lai, and C. Grebogi}}]{te2}
\bibinfo{author}{\bibfnamefont{W.~Z.} \bibnamefont{Zhang}},
\bibinfo{author}{\bibfnamefont{J.} \bibnamefont{Cheng}},
\bibinfo{author}{\bibfnamefont{J.~Y.} \bibnamefont{Liu}}
\bibnamefont{and} \bibinfo{author}{\bibfnamefont{L.}~\bibnamefont{Zhou}},
\bibinfo{journal}{Phys. Rev. A}
\textbf{\bibinfo{volume}{91}}, \bibinfo{pages}{063836}
(\bibinfo{year}{2015}).

\bibitem[{\citenamefont{Wang et~al.}(2014)\citenamefont{G. L. Wang, L. Huang, Y. C. Lai, and C. Grebogi}}]{Exx1}
\bibinfo{author}{\bibfnamefont{M.} \bibnamefont{Eichenfield}},
\bibinfo{author}{\bibfnamefont{R.} \bibnamefont{Camacho}},
\bibinfo{author}{\bibfnamefont{J.} \bibnamefont{Chan}},
\bibinfo{author}{\bibfnamefont{K.~J.} \bibnamefont{Vahala}},
\bibnamefont{and} \bibinfo{author}{\bibfnamefont{O.}~\bibnamefont{Painter}},
\bibinfo{journal}{Nature}
\textbf{\bibinfo{volume}{459}}, \bibinfo{pages}{550}
(\bibinfo{year}{2009}).

\bibitem[{\citenamefont{Wang et~al.}(2014)\citenamefont{G. L. Wang, L. Huang, Y. C. Lai, and C. Grebogi}}]{Exx2}
\bibinfo{author}{\bibfnamefont{M.} \bibnamefont{Eichenfield}},
\bibinfo{author}{\bibfnamefont{J.} \bibnamefont{Chan}},
\bibinfo{author}{\bibfnamefont{M.~R.} \bibnamefont{Camacho}},
\bibinfo{author}{\bibfnamefont{K.~J.} \bibnamefont{Vahala}},
\bibnamefont{and} \bibinfo{author}{\bibfnamefont{O.}~\bibnamefont{Painter}},
\bibinfo{journal}{Nature}
\textbf{\bibinfo{volume}{462}}, \bibinfo{pages}{78}
(\bibinfo{year}{2009}).

\bibitem[{\citenamefont{Wang et~al.}(2014)\citenamefont{Gerardo Adesso and Fabrizio Illuminat}}]{Na1}
\bibinfo{author}{\bibfnamefont{G.} \bibnamefont{Adesso}}
\bibnamefont{and} \bibinfo{author}{\bibfnamefont{F.}~\bibnamefont{Illuminat}},
\bibinfo{journal}{J. Phys. A: Math. Theor.}
\textbf{\bibinfo{volume}{40}}, \bibinfo{pages}{7821–-7880}
(\bibinfo{year}{2007}).

\bibitem[{\citenamefont{Wang et~al.}(2014)\citenamefont{C. Joshi J. Larson M. Jonson E. Andersson, and P. O hberg}}]{Na2}
\bibinfo{author}{\bibfnamefont{C.} \bibnamefont{Joshi}},
\bibinfo{author}{\bibfnamefont{J.} \bibnamefont{Larson}},
\bibinfo{author}{\bibfnamefont{M.} \bibnamefont{Jonson}},
\bibinfo{author}{\bibfnamefont{E.} \bibnamefont{Andersson}}
\bibnamefont{and} \bibinfo{author}{\bibfnamefont{P.~\"O.}~\bibnamefont{hberg}},
\bibinfo{journal}{Phys. Rev. A}
\textbf{\bibinfo{volume}{85}}, \bibinfo{pages}{033805-11}
(\bibinfo{year}{2012}).

\bibitem[{\citenamefont{Wang et~al.}(2014)\citenamefont{Yong Li Lian-Ao Wu and Z. D. Wang}}]{MR1}
\bibinfo{author}{\bibfnamefont{Y.} \bibnamefont{Li}},
\bibinfo{author}{\bibfnamefont{L.~A.} \bibnamefont{Wu}}
\bibnamefont{and} \bibinfo{author}{\bibfnamefont{Z.~D.}~\bibnamefont{Wang}},
\bibinfo{journal}{Phys. Rev. A}
\textbf{\bibinfo{volume}{83}}, \bibinfo{pages}{043804-5}
(\bibinfo{year}{2011}).

\bibitem[{\citenamefont{Wang et~al.}(2014)\citenamefont{Yong Li Lian-Ao Wu and Z. D. Wang}}]{MR3}
\bibinfo{author}{\bibfnamefont{P.~C.} \bibnamefont{Ma}},
\bibinfo{author}{\bibfnamefont{J.~Q.} \bibnamefont{Zhang}},
\bibinfo{author}{\bibfnamefont{Y.} \bibnamefont{Xiao}},
\bibinfo{author}{\bibfnamefont{M.} \bibnamefont{Feng}}
\bibnamefont{and} \bibinfo{author}{\bibfnamefont{Z.~M.}~\bibnamefont{Zhang}},
\bibinfo{journal}{Phys. Rev. A}
\textbf{\bibinfo{volume}{90}}, \bibinfo{pages}{043825-7}
(\bibinfo{year}{2014}).

\bibitem[{\citenamefont{Zhang et~al.}(2014)\citenamefont{Yong Li Lian-Ao Wu and Z. D. Wang}}]{MR2}
\bibinfo{author}{\bibfnamefont{J.~Q.} \bibnamefont{Zhang}},
\bibinfo{author}{\bibfnamefont{Y.} \bibnamefont{Li}}
\bibnamefont{and} \bibinfo{author}{\bibfnamefont{M.}~\bibnamefont{Feng}},
\bibinfo{journal}{J. Phys: Condens. Matter}
\textbf{\bibinfo{volume}{25}}, \bibinfo{pages}{142202-5}
(\bibinfo{year}{2013}).

\bibitem[{\citenamefont{Wang et~al.}(2014)\citenamefont{Yong Li Lian-Ao Wu and Z. D. Wang}}]{MR4}
\bibinfo{author}{\bibfnamefont{X.~Y.} \bibnamefont{L\"u}},
\bibinfo{author}{\bibfnamefont{J.~Q.} \bibnamefont{Liao}},
\bibinfo{author}{\bibfnamefont{L.} \bibnamefont{Tian}}
\bibnamefont{and} \bibinfo{author}{\bibfnamefont{F.}~\bibnamefont{Nori}},
\bibinfo{journal}{Phys. Rev. A}
\textbf{\bibinfo{volume}{91}}, \bibinfo{pages}{013834-8}
(\bibinfo{year}{2015}).

\end{thebibliography}

\end{document}